\documentclass[a4paper]{jpconf}
\usepackage{graphicx}
\usepackage[square,sort&compress]{natbib}
\usepackage{amsmath}
\usepackage{color}
\usepackage{soul}
\usepackage{times}
\usepackage{journals}

\def\thetav{\pmb{\theta}}

\begin{document}
\title{Current Issues in Asteroseismology}

\author{
   Michael Bazot$^{1}$,
   M\'ario J. P. F. G. Monteiro$^{1,2}$,
   Christian W. Straka$^{1}$}

\address{$^1$ Centro de Astrof\'{\i}sica da Universidade do Porto,
   Rua das Estrelas, 4150-762 Porto, Portugal}
\address{$^2$ Departamento de Matem\'atica Aplicada,
   Faculdade de Ci\^encia da Universidade do Porto, Portugal}


\begin{abstract}
 In this contribution we briefly review some of the current issues and promises for the future by asteroseismology.
 We are entering a new phase in this field driven by the wealth of data that has been collected and data that will soon be available for asteroseismology across the HR Diagram.
 Major difficulties in the descriptions of stellar interiors that arose in              
the second half of the 20th century may now be in part addressed and solved     
(this is the expectation!) by asteroseismology with unprecedented               
precision.
 In this contribution we list some of the key open questions in stellar physics, the seismic data we expect to collect in the near future, and some techniques that will provide the tools to connect data and models.

\end{abstract}

\section{Introduction}

 Stellar theory has been successful in explaining a wealth of observational data available for very different types of stars at various evolutionary stages.
 However, stellar physics was mainly confronted with explaining and predicting global stellar parameters such as luminosities and effective temperatures.
 So far, most of the strongest observational tests have come from studies of stellar clusters and helioseismology.
 These two methods rely on opposite philosophies.
 Cluster studies use large samples of stars, for which individual processes may be averaged out.
 The ongoing debate about the observational evidence of convective core overshoot in stars is just one example of the difficulties addressed by this method \cite{2001AJ....122.1602W}.
 Helioseismology, on the other hand, is concerned with the study of one star and, thus, provides information on a unique evolutionary state, and only for a specific stellar mass.
 Moreover, the Sun can be considered as a comparatively easier star to model: it does not have a central convective core and is only moderately evolved. 
 Therefore, the best solar models cannot be directly applied to other stars, whose interiors are determined by very different physical regimes.
 The consequence is a lack of direct constraints that can allow us to predict and explain the behaviour of these very different stars.
 
 Classical pulsators have in some cases allowed us to go beyond standard solar physics.
 These stars have opened the door to the opportunities offered by doing seismology across the HR diagram.
 Following the success of Helioseismology, the possibility of using oscillation data for most stars is now a reality following the observational evidence that almost all stars pulsate in some way \cite{Eyer08}.
 Asteroseismology has the potential, in particular, to fill in the apparent gap between helioseismology and the study of clusters.
 Detailed structural tests for individual stars with different masses and at different evolutionary stages will become possible with this technique.
 Thus, asteroseismology will provide us with much stronger constraints on the theory of stellar interiors.
 In particular, this technique will be fundamental in solving some of the long standing problems in stellar physics.
 Several physical processes have been included in models of stars and their interwoven effects on the overall structure cannot be discriminated using the currently available observational data.
 In Section~\ref{sec:issues}, below, some of these processes will be briefly discussed.

 One key aspect necessary to secure a strong impact of these new incoming data is the development of tools able to provide a direct connection between observations (oscillation mode parameters and global stellar parameters) and models. 
 These tools have been developed successfully in helioseismology.
 The drive is now strong towards developing equivalent inference methods for asteroseismology (e.g. \cite{Chaplin08}).
 This effort can take advantage of the solar experience but the techniques are different because of the limitations in the asteroseismic data.
 In particular, for stars other than the Sun, we are restricted to low degree modes.
 Nevertheless, these modes are able to constrain the physics of the models if adequate inference tools are applied. 
 Some of the aspects on what data we expect in the near future (Section~\ref{sec:data}) and how these data can be connected to the models (Sections~\ref{sec:link} \& \ref{sec:parameters}) are also discussed below.
 The paper ends with a brief comment on the (expected) future of Asteroseismology.

\section{Current issues in stellar physics}\label{sec:issues}

 Even though the Sun is by far the best studied star, solar modelling still faces problems.
 If the newly revised solar heavy element abundances are to be confirmed, solar models will clearly be at odds with the observational data from helioseismology \cite{Bahcall06}. 

 A number of physical processes that are not included in standard solar models might be relevant for solving remaining discrepancies between theory and observations.
 Among these are dynamical effects involving rotation and magnetic fields. 
 In particular, a mechanism to transport angular momentum from a presumed initially rapidly rotating to a uniformly rotating solar interior is needed.
 Stars in different evolutionary stages are expected to contain differentially rotating interiors, for which such mechanisms must also be included in the models. 
 As an example, the differential rotation is responsible for meridional circulation and shear-induced turbulence, which transport angular momentum and react back to the rotation \cite{2007EAS....26...49Z}.
 These mechanisms may modify the distribution of chemical elements and the magnetic fields (e.g. \cite{Straka07}).

 Another issue for which stellar models need better observational constraints, is convective core physics.
 All prescriptions for core overshooting or semiconvection have free parameters that must be calibrated through observations.
 Asteroseismology may provide additional constraints that would help to discriminate between different models/prescriptions. 
 Similarly, the physics of convective envelopes is not fully established.
 One example is the evidence given by the Sun on the inadequacies of the models to reproduce the transition layer at the base of the envelope. 
 In this case, the use of stars slightly different from the Sun may provide a clear indication on how the modelling should be improved.

 Seismic frequencies of high radial orders as well as the large frequency separations are sensitive to the structure of the outer envelope of stars. 
 Non-standard stellar processes relevant in this region include three dimensional effects of turbulent convection in the super-adiabatic layer, convective entrainment, rotation, magnetic fields and stellar winds. 
 The study of several stars covering a wide range in effective temperature is required in order to understand the origin and impact of the surface effects on the oscillation frequencies. 
 Seismic diagnostics complemented by 3D numerical simulations may be the only way forward to solve the questions about the near-surface physics of stars that remain unsolved, even for the Sun.

 These are some examples of the current issues in stellar physics that may be addressed with the use of precise stellar seismic data.

\section{Precise stellar data for Asteroseismology}\label{sec:data}

 In order to be able to confront the aforementioned issues in stellar physics with observational data it is necessary to have both accurate seismic data and global stellar parameters.
 The latter are important in helping asteroseismology to probe with high precision the physics of the stellar interior \cite{Creevey07,Cunha07}.
 The forthcoming GAIA mission alongside with high-precision spectroscopy and interferometry will allow the acquisition of accurate global parameters.

 In the following, we focus on the present and future developments of observational asteroseismology.
 Current space missions with programs dedicated to asteroseismology are MOST \cite{Walker08} and CoRoT \cite{Baglin06}.
 These should be followed by Kepler \cite{JCD08} and PLATO \cite{Catala05}.
 From ground, the emphasis is put on the already existing high-spectroscopy facilities and on the development of new projects such as SONG \cite{Grundahl08} and SIAMOIS \cite{Mosser08}. 
 A more detailed description of these missions and projects can be found throughout this volume.



\subsection{MOST}

 The MOST (Microvariability and Oscillations of STars) photometric satellite was launched in June 2003 and has already undertaken 64 primary campaigns and obtained observations of more than 850 secondary stars of which $\sim 180$ are variable.
 More than half of these variable stars pulsate, with the majority being B-type stars.
 MOST has detected $p$-modes in solar-type (e.g. red giants, yellow giants, $\alpha$ Cen A and B), in pre-main sequence, roAp and $\delta$ Scuti stars.
 An unanticipated discovery of MOST has been the detection of a large number of slowly pulsating B stars with variations that are characteristic of $g$-modes~\cite{2007arXiv0711.0706W}.


\subsection{CoRoT}

 The CoRoT \cite{Baglin06} photometric satellite was launched in December 2006 and has spent almost one year collecting data.
 The preliminary results confirm that the mission has achieved the project specifications and is able to produce data of unprecedented quality.
 The first release of data to the CoRoT community took place at the end of 2007.
 
 The primary goals of the CoRoT mission are to find planets and to use precise photometry for asteroseismology.
 A broad range of stellar types is expected to be observed from sun-like to very high mass stars.
 The various phases of stellar evolution will also be covered, ranging from young and main-sequence to evolved stars.

\subsection{Kepler}

 The Kepler mission is a NASA project to be launched in 2009 with the primary science objective of finding earth-like planets. 
 The Kepler data will also be used for a seismic study of the planet-hosting stars, because of the need to fully characterise the planetary systems being discovered.

 As a by-product, Kepler will also provide precise photometry for a large number of stars in its field of view \cite{JCD08}.
 The key scientific outcome of this additional program is the possibility of doing asteroseismology for a large number of solar-type stars.
 This wide coverage of stellar types and stellar evolution phases will open a new window on the understanding of the physics governing the evolution of the Sun and other stars. 
 Activity cycles \cite{Rempel08}, near-surface convection, and rotation \cite{Reese08} are among the issues of solar-type stars that will benefit from the constraints provided by the data to be collected by Kepler.

\subsection{Ground based facilities}

 For several years now, ground-based observations have been routinely performed to observe the variation of classical pulsators. 
 The situation has been quite different for sun-like stars. 
 It is well-known that the first breakthrough has come from the high-precision spectrograph CORALIE on the 1.5m Swiss-Telescope at La Silla \cite{Bouchy02}. 
 Since then, single-site observations, in particular those using the HARPS spectrograph, have led to several new $p$-mode identifications \cite{Bouchy05,Bazot07a}.

 In parallel to these single-site observations of solar-type stars, multi-site campaigns have been organized for other stars, namely $\alpha$~Cen A and B, using the spectrographs UVES and UCLES \cite{Butler04}, and Procyon \cite{Hekker08}. 
 The advantages of multi-site runs are obvious in terms of reducing the daily aliases. 
 However their organization can be difficult. 
 Firstly, the coordination of several telescopes depending on different organizations represents a delicate scheduling task. 
 Secondly, the instruments that are currently able to reach sufficiently high precisions do not allow a full coverage of the sky, and therefore of all relevant targets for ground-based seismology. 
 To overcome these difficulties the SONG project is being developed. 
 It consists in a 8-telescope network that would allow a full sky coverage in spectroscopy (see \cite{Grundahl08} for further details).
 
 There are other projects being developed/implemented with the primary goal of doing precise asteroseismology from the ground.
 The use of the South Pole is such an example (e.g. \cite{Mosser08})
or the development of a new generation of detectors for high precision spectroscopy for existing and future telescopes.

\section{The link between observations and models}\label{sec:link}

 In 1993, nearly ten years before the first reliable measurement of individual $p$-modes in $\alpha$~Cen A, Brown \& Christensen-Dalsgaard, in an article not so metaphorically entitled {\it How may seismological measurements constrain parameters of stellar structure?} \citep{Brown93}, proposed a prospective study that aimed at estimating the impact of asteroseismology on theoretical models.
 Today, thanks to the ever growing amount of available data, a not-so prospective answer to this question has to be sought. 
 This implies, in particular, developing the necessary methods for inference of the model parameters.

\begin{figure}
\begin{center}
\includegraphics[width=0.75\textwidth]{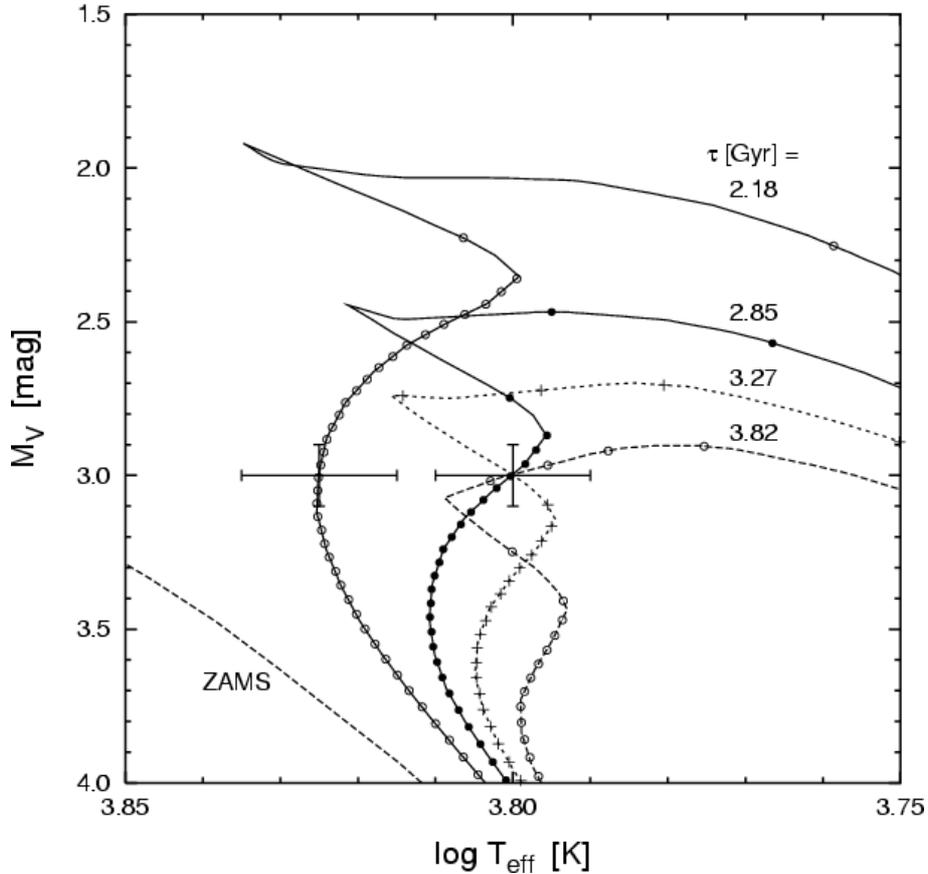}
\end{center}
\caption{\label{fig_jorgensen}
 HR diagram showing the location of two hypothetical observations at $(\log T_{\rm eff},M_V)=(3.825,~3.0)$ and (3.800, 3.0). 
 The zero-age main sequence (ZAMS) and selected isochrones for ages $\tau = 2.8, 2.85, 3.27 $ and 3.82~Gyr, for $\log(Z/Z_\odot) = {-}0.2$ are also shown. 
 The symbols along the isochrones show where the initial mass is a multiple of $0.01~{\cal M}_\odot$.
 (From J\o rgensen et al. 2005, \cite{Jorgensen05}).}
\end{figure}

 Choosing a method to estimate stellar parameters is in itself a non-trivial problem. 
 An illustration of this situation is given in Fig.~\ref{fig_jorgensen}. 
 The isochrones show a typical non-linear behaviour in the HR diagram, undergoing a hook when convective cores start to develop in the innermost regions of the corresponding stars. The rightmost set of observables in Fig.~\ref{fig_jorgensen} can be reproduced by several isochrones. 

 The number of free parameters necessary to describe a stellar model is significant.
 If one considers a so-called ``standard model'' for a sun-like star, the minimum number of input parameters is given by the mass, the age and the initial chemical composition (i.e. the initial helium and metal abundances) of the star.
 In the framework of the mixing-length theory, which describes convection, the mixing-length parameter should be included, which adds up to five parameters to be taken into account. 
 It is therefore of no surprise that the solution to this inverse problem can be degenerate.
 The situation is even more complex if we consider that many non-standard effects could be added in the stellar models which usually introduce additional free parameters (e.g., turbulence, rotational mixing, accretion, winds \cite{Straka06,Eggenberger08,Bazot04,Noels08}).
 Although the relevant physical parameters may change from one type of star to the other, this statement remains valid.

%
%
%

 Adding seismic constraints to the set of observational data certainly reduces the number of possible solutions, but the general difficulty of searching the multi-dimensional space of stellar input parameters remains.
 It is thus worth exploring the different available methods to estimate these parameters and their respective advantages and limitations.


\subsection{Stellar parameter search using optimization}




\subsubsection{Scanning pre-computed grids}

 One of the first methods developed to search for the 'best-fitting' model using asteroseismic constraints consists in systematically scanning pre-computed models to find matches to the oscillation frequencies.
 In some cases, very dense grids of stellar models, parametrized by mass, age and composition, have been used \cite{Guenther04}. 
 In this work, the authors compared each model in the grid with the seismic constraints, rejecting or including models based on the minimization of a $\chi^2$ merit function. 
 
 The main advantage of this method is that once a sufficiently large model
library has been built it can be quickly searched whenever new
asteroseismic data becomes available.
 On the other hand, once the number of parameters is extended, e.g. to include non-standard physics, the storage and computational requirements become increasingly demanding.

\subsubsection{Non-linear least square methods}


 Several methods for solving non-linear least square problems exist. 
 These consist in minimizing a $\chi^2 (\thetav)$ merit function, where $\thetav$ is an array containing the input parameters to the model, by using the estimates of the gradient. 
 In this way, a systematic scan in the space of parameters can be avoided. 
 The best-known, and perhaps the most widely used, method is the Levenberg-Marquardt algorithm. 
 It interweaves the principles of conjugate gradient and variable metric methods \cite{Press92}. 
 This approach has the convenient proprieties of being relatively fast and robust against variations of the initial guess of the parameters. 
 Its major limitation is the impossibility of identifying multiple solutions. 

 The Levenberg-Marquardt algorithm has been used with success in the case of $\alpha$~Cen A \cite{Miglio05}.
 However, it remains to be seen how efficient it would be in the study of higher-mass stars or stars in latter evolutionary stages.



\subsubsection{Global optimization}

\begin{figure}
\begin{center}
\includegraphics[width=0.75\textwidth]{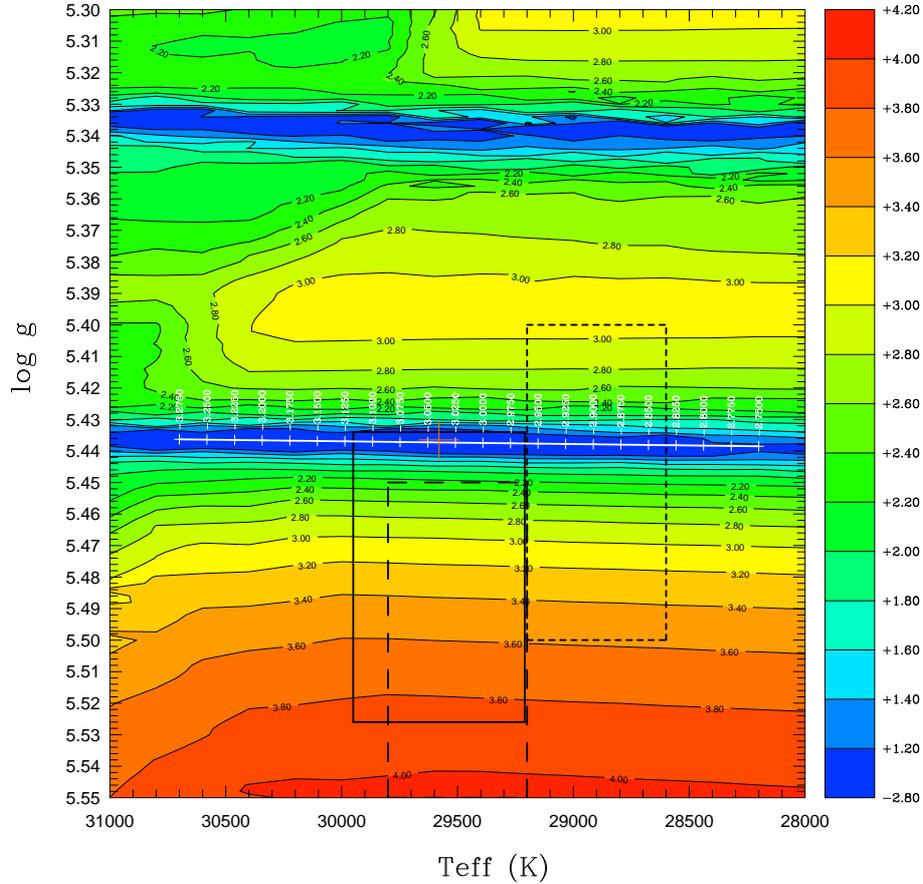}
\end{center}
\caption{\label{fig_charpinet}
 Slice of the ``projected'' $\overline {{\chi }^{2}}$-function (in logarithmic units) along the $\log g-T_{{\rm eff}}$ plane. 
 The labelled white line -- positioned along the valley (blue region) of low values of $\overline {{\chi }^{2}}$ -- indicates the location of the local minimum of $\overline {{\chi }^{2}}$ found with different values of the hydrogen-mass fraction $\log q({\rm H})$ (this parameter has been varied between -2.0 and -5.0 in steps of 0.025). 
 The rectangles represent different spectroscopic estimates (and associated uncertainties) for the atmospheric parameters of Feige~48.
 (From Charpinet et al. 2005, \cite{Charpinet05b})}
\end{figure}

 Alternative optimization algorithms are necessary if the function $\chi^2(\thetav)$ has multiple local minima (or even if it is non-differentiable).
 One of the possible solutions to this problem is the use of genetic algorithms \cite{Charbonneau95,Brassard01,Metcalfe03}.
 These are global optimization methods that do not use gradients.
 They rely on a stochastic sampling of the parameter space and are mainly used to avoid trapping in one single minimum.
 So far, these methods have been applied to the case of white dwarfs and SdB stars \cite{Charpinet05a,Charpinet05b,Randall08}.
 An example of the results produced by such a technique is given in Fig.~\ref{fig_charpinet}.
 It represents the mapping in a $T_{\mathrm{eff}}-\log g$ diagram of an average function $\overline{\chi^2}(\thetav)$ (constructed to give the best fit possible to the observed oscillation periods), for the SdB~Feige~48.
 The genetic algorithm has made possible the exploration of a physically relevant region of the space of parameters, necessary to understand the complex structure of $\overline{\chi^2}(\thetav)$.


 Methods based on stochastic sampling are usually much more time-consuming than local optimization methods. 
 For this reason, the asteroseismology of sub and white dwarfs has been a privileged field for their experimentation.
 Modeling of the stellar structure for these stars can be done with static codes
, which are much faster than their evolutionary counterparts used for modeling main sequence stars.
 Adapting these methods to other potential seismic targets represents the next step towards the application of global optimization to asteroseismology across the HR diagram.

\subsection{The Bayesian approach}

 Optimization is mainly concerned with the identification of a best model. 
 Even though the problem of multiple minima can be bypassed by the use of genetic algorithms, these do not indicate the probability associated with each possible solution.
 Consequently the genetic algorithms do not provide a confidence interval for the inferred parameters.

An alternative approach consists in describing the parameters using their posterior probability density function, conditional to the observations and to the assumptions made in the theoretical model, i.e. adopting a Bayesian point of view.
 So far, this approach has been used mainly to determine stellar ages from isochrones \cite{Pont04,Jorgensen05} and for more general parameter estimations concerning planet-hosting stars \cite{Takeda07}.
This method has several advantages. Firstly, it allows one to formalize the prior information existing on the parameters, via Bayes's theorem. Secondly, it is possible to use the computed probability density jointly with statistical inference to estimate the stellar parameters.
 The main drawback of the Bayesian approach is the computational requirements.
 The development of Bayesian estimation techniques in the field of asteroseismology is work in progress. 
It should be noted that Monte Carlo Markov Chain methods, which allow
one to scan the space of parameters using the power of stochastic
sampling and also relies on the Bayesian methodology, are currently being tested for seismic targets \cite{Bazot08b}.

\subsection{Estimation of confidence intervals}

 In order to qualify the solution that best represents the data it is necessary to determine confidence intervals for the estimated parameters. 
 We briefly mention here two approaches. The first one is the singular value decomposition (SVD) \cite{Brown94,Creevey07}. It consists in linearizing the observables as functions of the parameters in the vicinity of an identified solution of the optimization problem. Another approach consists in using Monte Carlo methods (see e.g. \cite{Jorgensen05}). These are robust for error estimation \cite{Press92} in particular when dealing with non-linear models. So far, Monte Carlo estimates for errors have not been extensively used, the most notable attempt concerns the Sun \cite{Bahcall06}. The main reason for this lack of applications in stellar physics are the heavy computational requirements imposed by random exploration methods. Nevertheless, some implementations are starting to become available for other stars \cite{Bazot08b}. 
 

\section{How efficient is asteroseismology?}\label{sec:parameters}

 Before concluding, we must also address the question of how effectively the seismic data can constrain the stellar parameters. 
 A qualitative evaluation of seismic studies of both white dwarfs and sun-like stars, confirms the intricate complementarity between seismic and spectro-photometric data in constraining the models. 
 Figure~\ref{fig_charpinet} shows a $\chi^2$ map projected in a $T_{\mathrm{eff}}-\log g$ diagram.
 The most striking feature is the horizontal (blue) valley, appearing at the middle of the figure, corresponding to the locus of solutions able to reproduce the seismic data.
 The improvement due to seismic data (blue valley), when comparing with the constraints resulting solely from spectroscopy (black rectangles), is striking.
 But the need to have precise spectrometric measurements, to complement the seismic data, is clear if all parameters are to be estimated with high accuracy. 
 Similar situations are found for sun-like stars such as $\alpha$ Cen A, $\mu$ Ara and $\beta$ Vir \cite{Miglio05,Bazot05,Eggenberger06}. 
 Therefore, any asteroseismic campaign should be complemented by a dedicated program aiming at obtaining high-precision spectro-photometric constraints for the target stars.

 To illustrate the impact of having seismic data we would like to cite some quantitative uncertainties derived from seismic studies, along with some theoretical estimates. 
 Using linearizing methods, Charpinet et al. \cite{Charpinet05a,Charpinet05b} obtained precisions of $\sim2\%$ on the mass of the SdBs Feige~48 and PG~1219+534, and about $\sim$3-3.5\% on the mass of their hydrogen-rich envelopes.
 Theoretical predictions for solar-type stars, based on the combination of precise seismic and interferometric measurements of the radius (below $\sim2\%$), indicate that an independent measurement of the stellar mass can be provided with an uncertainty below 3\% (see \cite{Creevey07} for details).  

 It is equally important to establish how each observational constraint may provide additional information on a specific stellar parameter in the modelling.
 Again, the evidence is that some parameters, such as, e.g., the mixing-length parameter, require seismic data to be determined with some accuracy.
 If a precise determination of the stellar mass is available, either directly or by using an interferometric measurement of the radius, it was shown by \cite{Monteiro02,Creevey07} that the seismic data can be used to probe the physics of the stellar interior (convection, diffusion or near-surface physics).
 Consequently, by selecting individual stars across the HR diagram for which seismic and non-seismic observations are available with high precision, asteroseismology will be able to constrain the physics going into stellar models. 
 Detailed studies of $\alpha$~Centauri, in particular, have already confirmed that the techniques needed to test the physics of stellar models are able to achieve very promising results.
 
 Diagnostic tools to extract from the oscillation data, specific indications on the stellar interior have also been developed for the Sun and extended to other stars with success.
 These methods can provide very precise tests of the physics, namely on the Helium ionization zones \cite{Monteiro98,Monteiro05,Houdek07}, the base of a convective envelope \cite{Monteiro00} or the super-adiabatic layer of solar-type stars \cite{Straka06}, to list a few.
 For further details we refer the reader to the literature \cite[e.g,][]{Cunha07,Straka07,Eggenberger08}.

\section{A new era for Asteroseismology}

 As it has been so well put by Laurent Eyer \cite{Eyer08}, {\em now is the time to be an Asteroseismologist}!
 After reviewing the list of contributions presented at the HELAS II conference in G\"ottingen the authors can only agree that the wealth of results that are starting to appear, as a consequence of the new data made available by ground based and space missions, is marking the beginning of a new era for stellar physics.

 The forthcoming years in this field are expected to see a major breakthrough, as the ``wave'' of new data inundating the community will lead undoubtedly to a much deeper understanding of the physics of stars.
 It may be (hopefully!) a big ``wave'', but the contributions presented at the HELAS II conference clearly show that the community is prepared to surf this wave (as the Sun has proved to be a good training ground).
 We look forward to the bright future of asteroseismology!

\ack

 The authors would like to thank M.~Cunha for her suggestions on an earlier version of the manuscript and S.~Charpinet and L.~Lindegren for the figures.
 This work was supported in part by the European Helio- and Asteroseismology
Network (HELAS), a major international collaboration funded by the
European Commission's Sixth Framework Programme.
 The authors acknowledge the support by FCT and FEDER (POCI2010)
through projects {\small POCI/CTE-AST/57610/2004} and {\small
POCI/V.5/B0094/2005}.

\bibliography{goettingen_ref}

\end{document}